\documentclass[]{aastex62}

\received{April 5, 2019}
\revised{May 15, 2019}
\accepted{May 28, 2019}
\submitjournal{ApJ}

\shortauthors{Kidder et al.}

\usepackage[section]{placeins}

\begin{document}

\title{Radial Velocities of Low-mass Candidate TWA Members}

\correspondingauthor{Benjamin Kidder}
\email{btkidder@astro.as.utexas.edu}

\author{Benjamin Kidder}
\affil{University of Texas at Austin}

\author{Gregory Mace}
\affiliation{University of Texas at Austin}

\author{Kimberly Sokal}
\affiliation{University of Texas at Austin}

\author{Ricardo Lopez}
\affiliation{University of Texas at Austin}

\author{Daniel Jaffe}
\affiliation{University of Texas at Austin}

\begin{abstract}

Nearby young moving groups provide unique samples of similar age stars for testing the evolution of physical properties.
Incomplete and/or incorrect group membership classifications reduce the usefulness of the group, which we assume to be coeval.
With near-infrared spectra of two candidate members of the TW Hya Association, 2MASS~J12354615$-$4115531 (TWA~46) and 2MASS J12371238$-$4021480 (TWA~47), we test their membership by adding radial velocity measurements to the literature. We find that 2MASS~J12354615$-$4115531 is a close spectroscopic binary system with a center-of-mass radial velocity of -6.5$\pm$3.9 km s$^{-1}$. 
This radial velocity and a {\it Gaia} parallax produces a TWA membership probability of 41.9$\%$ using the Banyan $\Sigma$ tool for 2MASS~J12354615$-$4115531. The spectrum of 2MASS J12371238$-$4021480 shows that it appears to be a single star with a radial velocity consistent with the TW Hya Association and a membership probability of 99.5$\%$. 
The reduced probability of TWA~46 as a true member of TWA highlights the importance of high-resolution, near-infrared spectra in validating low-mass moving group members.

\end{abstract}

\keywords{TW Hya Association, Young Stellar Objects, Young Moving Groups, Spectroscopic Binaries, M dwarfs}

\section{Introduction} \label{sec:intro}

Stars with intermediate ages (between a few to 750~Myr) are difficult to separate from the much older field population without thorough investigation of spectral indicators of youth. 
However, young moving groups (YMGs) of spatially and kinematically related stars allow the use of well calibrated FGK-type stars to age date the entire group down to the substellar population \citep{Zuckerman2004,Torres2008,gagne2018b}. 
The TW Hydra association (TWA), with an age of $\sim$10~Myr \citep{bell2015}, is the youngest of these YMGs within a hundred parsecs. 
Recent analysis by \citet{sokal2018} of TW Hydra, the namesake of the TW Hydra Association, found an age of $\sim$8~Myr.

Nearby young stellar objects are frequently proposed as candidate YMG members based on their proper motion. Converting these candidate YMG members into validated members requires verification of their spatial and kinematic commonality with defining moving group members \citep{gagne2018b}. 
Only with accurate confirmation of YMG membership can they be used to study stellar evolution or make assumptions about the entire group's characteristics.
\cite{gagne2017} presents updated membership information for TWA, including 30 bona fide members and 13 high-likelihood members that have all been assigned TWA designations but still require follow-up observations to confirm membership beyond a reasonable doubt. 
One such high-likelihood member is 2MASS J12354615$-$4115531 (hereafter 2MJ1235), which was found to show signs of youth, including strong H$\alpha$ and X-ray emission \citep{riaz2006}, prior to being identified as a candidate member of TWA. 
2MJ1235 was classified as a TWA member by \cite{donaldson2016} based on an isochronal age estimate of 23$\pm$12 Myr and an astrometric solution calculated using parallax and proper motion, but not radial velocity (RV). 
In this paper we use a high-resolution, near-infrared (IR) spectrum of 2MJ1235 to show that it is a spectroscopic binary, and provide updated information on the binary pair's questionable TWA membership. 
Additionally, we report an RV for TWA candidate member 2MASS J12371238$-$4021480 (hereafter 2MJ1237) that is more consistent with TWA membership than that given by \citet{riedel2016}. They identified 2MJ1237 as a potential TWA member based on its RV and proper motion, in addition to strong H$\alpha$ emission and the presence of lithium. \cite{gagne2017} gave 2MJ1237 the designation of TWA~47 based on the findings of \citet{riedel2016}.

\section{Observations}
We obtained high-resolution near-IR spectra of 2MJ1235 and 2MJ1237 using the Immersion Grating Infrared Spectrometer (IGRINS) at Gemini South Observatory \citep{mace2018spie}. IGRINS covers all of the H and K photometric bands in a single exposure at a resolution of $\sim$45,000 \citep{yuk2010,park2014}. The spectrum is cross-dispersed into $\sim$40 overlapping echelle orders. To achieve such broad wavelength coverage at high-resolution, IGRINS uses a silicon immersion grating \citep{gully2012}, which increases the resolution of the grating and the dispersion angle by a factor of the index-of-refraction of silicon (n$=$3.4). 

IGRINS observations of several other objects are used in this work as templates for determining RVs and flux ratios. 
An observation summary is provided in Table \ref{obs_params} for the two candidate members of TWA and several other objects used in this analysis. All observations were reduced using the IGRINS pipeline package version 2.2 \citep{lee2016}. The wavelength solution was determined using telluric absorption and OH emission lines. OH emission lines were removed prior to spectral extraction through A-B pair subtraction, and the remaining telluric absorption features were removed by dividing the spectrum by that of an A0V telluric standard. All A0V spectra used for telluric correction in this work were observed within $\sim$1 hour of the target spectra, with the same telescope and instrument configuration, and at a similar airmass. The final product of the IGRINS pipeline is a one-dimensional, telluric-corrected spectrum. All spectra used in our analysis were barycenter velocity corrected prior to cross correlation employing corrections derived using ZBARYCORR \citep{wright2014}. 

\section{Analysis \label{analysis}}
The primary quantities derived for 2MJ1235 and 2MJ1237 are the RVs, which we use in conjunction with the BANYAN $\Sigma$ tool \citep{gagne2018} to test moving group membership. Initial inspection of 2MJ1235 revealed doubled spectral lines, as shown in Figure~\ref{binary_fit}. Therefore, we measured the RV of both components of 2MJ1235AB, and the flux ratio, in order to determine a center-of-mass RV for the pair.

\subsection{2MJ1235AB Component Radial Velocities}

\begin{figure}[ht]
\centering
\includegraphics[width=3.5in]{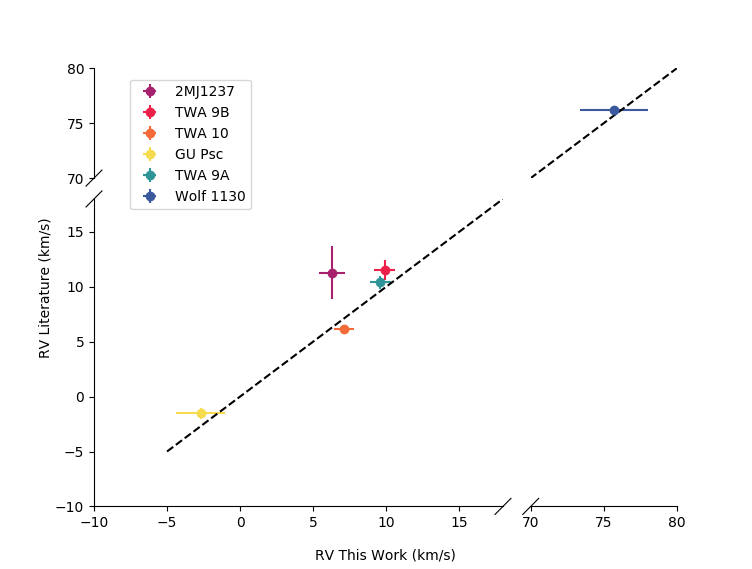}
\caption{ Comparison of RV measurements from the literature (see Table \ref{obs_params} for references) with RVs measured by cross-correlating IGRINS spectra of objects with well known RVs.\label{rv_compare}}
\end{figure}

Radial velocities for both components of 2MJ1235 were measured by cross correlating with IGRINS spectra of the objects listed in Table \ref{obs_params}. 
In order to find the best template spectrum for determining the RVs of 2MJ1235AB we cross correlated it with IGRINS spectra of bona fide TWA members.
We selected TWA 10 as the best RV template for both components based on the height of the cross-correlation peaks and the accuracy of its own RV measurement in the literature  \citep{elliott2014}. 
To determine the RVs of both components of 2MJ1235AB we selected 31 IGRINS orders across the H- and K-bands that had the least residual telluric contamination. 
We cross-correlated each of the individual selected IGRINS orders of TWA 10 and 2MJ1235AB and determined the locations of the two maxima for the 31 resulting cross-correlation functions. 
The component RVs and associated errors are the mean and 1~$\sigma$ uncertainty for all 31 IGRINS orders.
The absolute RVs for the components of 2MJ1235AB were determined by subtracting the literature RV of TWA 10 and propagating the uncertainties. 
We also designate the red-shifted component in 2MJ1235AB to be the primary based on the fact that its absorption lines are consistently deeper than those of its companion. 
We measure RVs of 56.6$\pm$0.5 km s$^{-1}$ for 2MJ1235A, and -79.9$\pm$0.8 km s$^{-1}$ for 2MJ1235B.
A comparison of our measured RVs for all target and template objects in this paper is given in Table~\ref{obs_params}, and a comparison between our measured RVs and those taken from the literature is shown in Figure \ref{rv_compare}.

\begin{deluxetable*}{cccrrclc}[ht]
\tablecaption{Observation log and stellar parameters. \label{obs_params}}
\tablecolumns{8}
\tablenum{1}
\tablewidth{0pt}
\tablehead{
\colhead{Object} &
\colhead{UT Obs Date} &
\colhead{MJD} &
\colhead{Literature RV} & \colhead{This Work RV} & \colhead{BVC} & \colhead{SpT} & \colhead{Ref.\tablenotemark{a}} \\
\colhead{} & \colhead{(YYYY-mm-dd)} & \colhead{(d)} &
\colhead{(km s$^{-1}$)} & \colhead{(km s$^{-1}$)} & \colhead{(km s$^{-1}$)} & \colhead{} & \colhead{}
}
\startdata
2MJ1235 & 2018-04-24 & 58232.0861 & & -6.47$\pm$3.9   & -2.73  & M3.4 + M3.4 & This work     \\
2MJ1237					& 2018-04-27 & 58235.1073 & 11.3$\pm$2.4    & 6.3$\pm$0.9 & -4.23  & M2.5 Ve & 1, 2  \\
TWA 9A 					& 2018-05-08 & 58246.0706   & 10.4$\pm$0.6  & 9.6$\pm$0.7 &-13.09 & K6  & 3, 4  \\
TWA 9B                  & 2018-04-25 & 58233.0257   & 11.5$\pm$0.9  & 9.9$\pm$0.7 &-7.96  & M3.4    & 3, 4  \\
TWA 10 					& 2018-04-28 & 58236.0241   & 6.2$\pm$0.3   &  &	-4.80  & M2 Ve   & 3, 5  \\
GU Psc                  & 2014-11-23 & 56984.0602   & -1.5$\pm$0.5  & -2.7$\pm$1.7 &-18.37 & M3      & 6, 7  \\
Wolf 1130               & 2015-08-05 & 57239.2435   & 76.18$\pm$0.16         & 75.7$\pm$2.3 &3.91   & sdM1.5    & 8, 9 
\enddata
\tablenotetext{a}{Reference for RV and SpT}
\tablecomments{TWA 10 has no measured RV in this work because we use its literature RV to set our absolute RV zero-point.}
\tablereferences{(1)\cite{riedel2016}; (2)\cite{gagne2017}; (3)\cite{elliott2016}; (4)\cite{herczeg2014}; (5)\cite{torres2006}; (6)\cite{malo2014}; (7)\cite{riaz2006}; (8)\cite{mace2018}; (9)\cite{gizis1997}}
\end{deluxetable*}

\subsection{2MJ1235AB Flux Ratio and Center-of-Mass Velocity}
\begin{figure}[ht]
\centering
\includegraphics[width=6.2in]{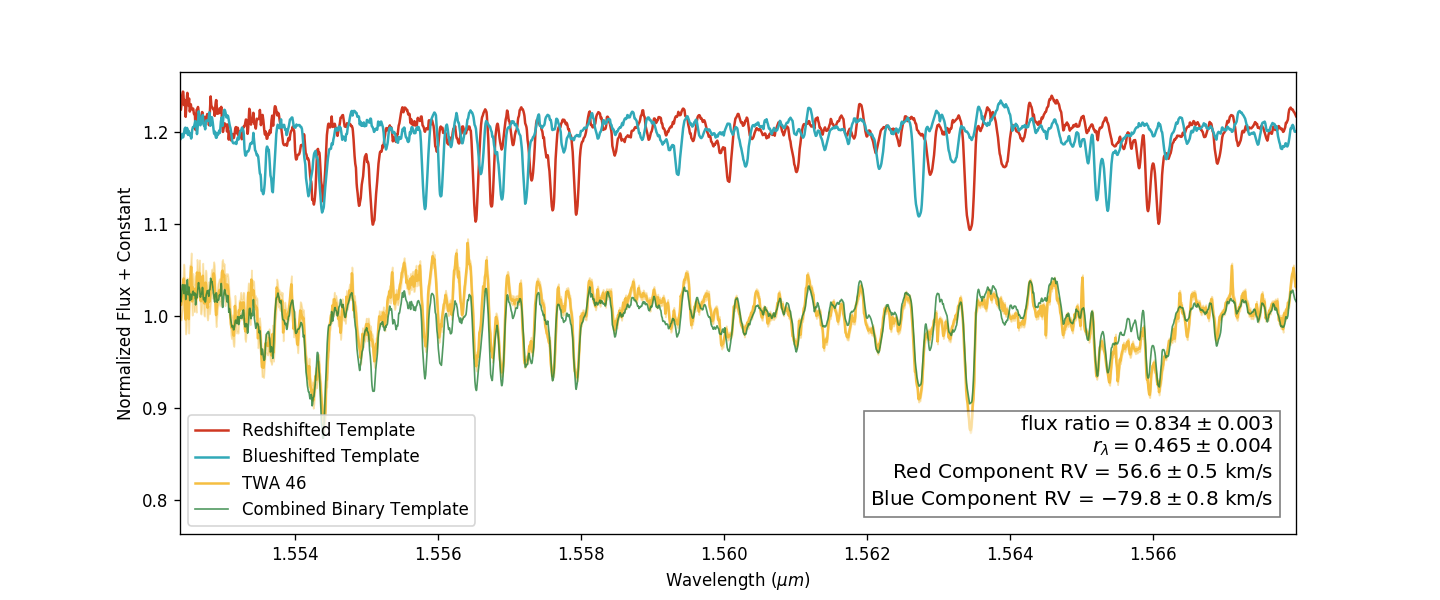}
\includegraphics[width=6.2in]{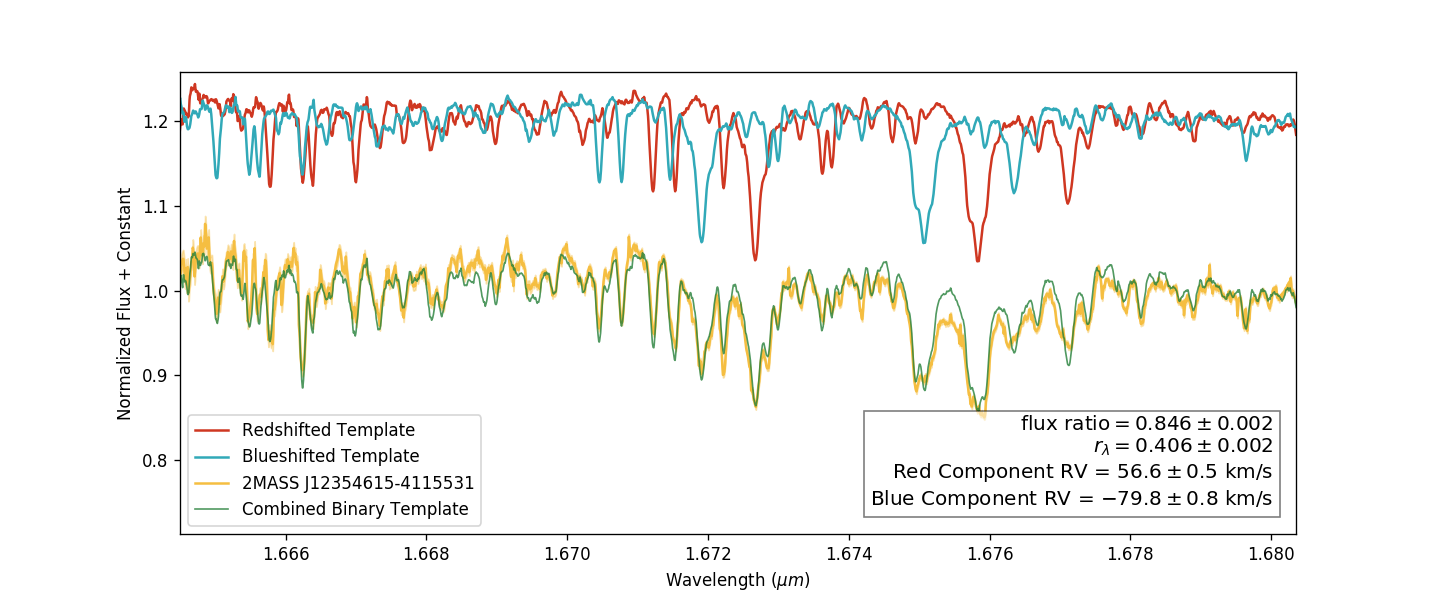}
\includegraphics[width=6.2in]{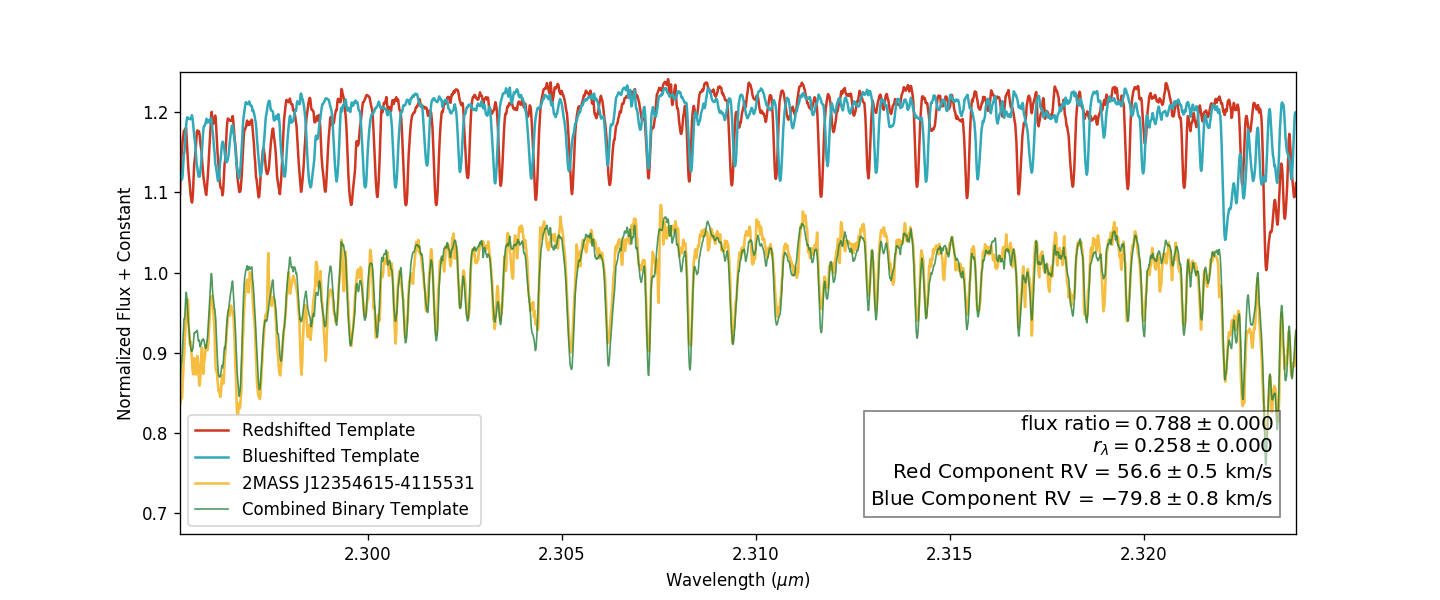}
\caption{Binary template fits to three orders of the IGRINS spectrum of 2MJ1235AB using TWA 9B as the template for both the primary and secondary component. Each figure shows the spectrum of TWA 9B shifted to the wavelength of the primary and secondary components of 2MJ1235AB and adjusted to the flux ratio found by the fit. The figures also show the spectrum of 2MJ1235AB and the combined binary template after the duplicated spectra of TWA 9B have been shifted, fit for flux-ratio, veiling and continuum offset, and combined into a binary template. The results of fitting for flux ratio and veiling, which were measured independently for each order, are displayed on each of the images above.\label{binary_fit}}
\end{figure}

Without multi-epoch spectra of 2MJ1235AB to provide a full orbital solution and center-of-mass velocity, we use the flux ratios in the double-lined IGRINS spectrum to derive corresponding K-band magnitudes from which to estimate component masses.
Composite binary spectral templates for each of the 31 selected IGRINS orders were created by offsetting the single-object templates (Table~\ref{obs_params}) to the RVs determined in Section 3.1. The combined template spectrum was then optimized by fitting the flux ratio using the Markov chain Monte Carlo (MCMC) ensemble sampler $\mathrm{emcee}$ \citep{mackey2013}. Veiling and intrinsic continuum-offset parameters were applied to the combined binary template when fitting each order. In order to determine the best fitting template for each component of the binary, we performed this fitting routine using every possible combination of the 23 TWA members for which we have acquired IGRINS spectra. This brute-force template selection method, while not computationally efficient, was practical given the limited size of our template library. 
The best-fitting template for both 2MJ1235A and 2MJ1235B was TWA 9B \citep[SpT M3.4V,][]{herczeg2014}. Figure \ref{binary_fit} shows the fit of the binary template to 2MJ1235AB for three orders of the IGRINS spectrum. 
Fitting for the flux-ratio, using TWA 9B as the template, resulted in a secondary-to-primary flux ratio of 0.86$\pm 0.08$. 
The near-infrared flux ratio can be used as a proxy for the mass ratio \citep[e.g.][]{kraus2008}, and absolute K-band magnitudes provide direct mass estimates \citep{mann2019}.
Based on the flux ratio, we compute a center of mass velocity for the system of -6.47$\pm$3.9 km s$^{-1}$.
The BANYAN $\Sigma$ tool \citep{gagne2018} then produces a 41.9$\%$ TWA membership probability for 2MJ1235AB.

\subsection{Age, Mass and Orbital Parameter Estimates}

\begin{figure}[ht]
\centering
\includegraphics[width=4.5in]{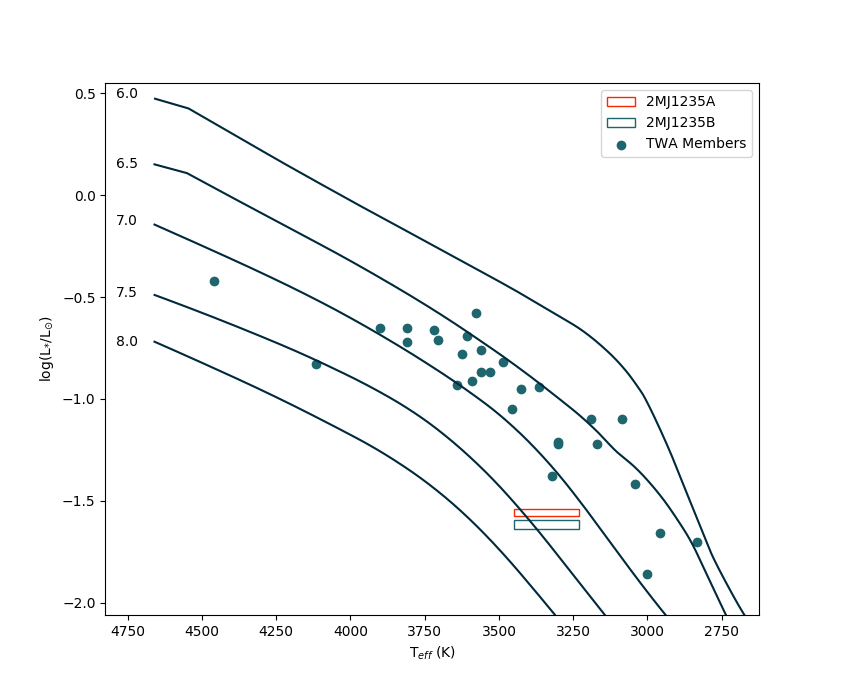}
\caption{H-R diagram of known TWA members with T$_{\mathrm{eff}}$ and luminosities from \cite{herczeg2014} and the parameter space that 2MJ1235A and B occupy. We adopted T$_{\mathrm{eff}}$ for 2MJ1235A and B from the spectral type relationship of \cite{herczeg2014}, while the calculated luminosity is based on our measurement of the flux ratio and the parallax from \textit{Gaia} \citep{brown2018}. We plot age isochrones from \cite{baraffe2015} at intervals of 0.5 dex in log(yr). \label{HR_diagram}}

\end{figure}

The measured flux ratio for the system, in combination with photometric and parallax information, allows us to place limits on the orbital separation and period of 2MJ1235AB. 
Having measured the flux ratio for the binary pair, we used the H- and K-band magnitude from 2MASS \citep{skrutskie2006} and parallax from {\it Gaia} \citep{brown2018} to determine absolute magnitudes for the system components (H$_{A,B}=$~6.37, 6.53 mag; K$_{A,B}=$~6.13, 6.29 mag). 
The absolute magnitudes for the individual components, along with the bolometric correction from \cite{mann2015}, allow us to place 2MJ1235A and 2MJ1235B on the Hertzsprung–Russell (H-R) diagram. We adopt a temperature of 3340$\pm$110K for both members of 2MJ1235AB based on the best-fitting template (TWA 9B), the temperature sequence from \cite{herczeg2014}, and an assumed uncertainty of 0.5 spectral types. 
A comparison of 2MJ1235A and 2MJ1235B on the H-R diagram to the evolutionary tracks from \cite{baraffe2015} produce an age range of 12-50~Myr (Figure~\ref{HR_diagram}).  This age range extends 14~Myr above the age determined by \cite{donaldson2016} (23$\pm$12 Myr), which was calculated under the assumption that the newly identified binary was a single star. Given the accepted age of TWA, $\sim$10~Myr \citep{bell2015}, 2MJ1235AB appears to be an age outlier.

To estimate the component masses for 2MJ1235A and B we can employ the empirical mass-luminosity relationship in \cite{mann2019} along with absolute K-band magnitudes. 
However, the mass-luminosity relationship outlined by \cite{mann2019} is recommended for M-dwarfs older than 100 Myr, and the derived masses ($\sim$0.4~M$_{\odot}$) should be considered upper limits. 
Evolutionary tracks from \cite{baraffe2015} suggest that a 0.3-0.4 M$_{\odot}$ star will reduce its radius by half when it ages from 8-50 Myr. If 2MJ1235AB is in fact a TWA member at 8 Myr \citep{sokal2018} it may be as many as 5 times over-luminous for the \cite{mann2019} relationships. Taking the age differences into account we find lower mass limits of $\sim$0.2~M$_{\odot}$ for both 2MJ1235A and B.


2MJ1235AB was observed in a series of six 300s exposures, which we initially reduced all together to maximize signal-to-noise in the prior analysis. 
However, we also reduced each of the three AB pairs separately and found that the velocity separation between the two components of the binary increased by $\sim$1.5 km s$^{-1}$ between the first and last AB pair, which were taken $\sim$23 minutes apart. We estimate that the error for the individual RV measurements for the separated AB pairs is $\pm$0.7km/s. While the statistical significance of this acceleration is tenuous, it is also small enough to imply a lower limit on the orbital period of several hours.
The inclination for the system and orbital phase of our observation are unknown, so our measured velocity separation for the pair provides only a lower limit on the orbital velocity semi-amplitudes. Along with our estimated upper limits on the component masses, the upper bound on the orbital separation is 5.8$\times$10$^6$~km and the maximum orbital period is $\sim$3.2~days. A summary of the properties of 2MJ12354615 is provided in Table \ref{2MJ1235_params}.

\begin{deluxetable*}{cccrrclc}[ht]
\tablecaption{Properties of 2MASS~J12354615-4115531~AB \label{2MJ1235_params}}
\tablecolumns{2}
\tablenum{2}
\tablewidth{0pt}
\tablehead{
}
\startdata
F$_{2}$/F$_{1}$ & 0.86$\pm$0.08\\
2MASS H (mag) & 9.468$\pm$0.023 \\
2MASS K (mag) & 9.232$\pm$0.023\\
Gaia DR2 parallax (mas) & 17.59$\pm$0.17\\
Age\tablenotemark{a} (Myr) & 12-50 \\
Mass (M$_{\odot}$) & 0.2-0.4\\
log~$L_{A,B}/L_{\odot}$ & -1.55$\pm0.02$, -1.61$\pm0.02$\\
Orbital separation (km) & $<$5.8$\times 10^6$\\
Orbital period (days) & $<$3.2
\enddata
\tablenotetext{a}{\citet{baraffe2015} evolutionary tracks for luminosity and T$_{\mathrm{eff}}$}
\end{deluxetable*}

With a single epoch of observation, we have identified excellent template matches for both components of 2MJ1235AB, and estimate masses based on template fitting. However, multiple epochs of observations of 2MJ1235AB at different phases would allow for the spectra of the two objects to be fully separated and for stellar parameters of the two objects to be measured independently of one another. Measuring parameters for the individual components for the system may allow us to measure the system's properties more precisely. 

\subsection{2MASS~J1237 (TWA 47)}
The second candidate member of TWA that we observed was 2MJ1237. 
We performed some of the same analysis on 2MJ1237 as 2MJ1235AB and found that TWA 9A \citep[SpT K6,][]{herczeg2014} was the best overall match.
However, we chose TWA 10 as the RV template because of its more precise literature RV \citep{elliott2016} and higher signal-to-noise ratio IGRINS spectrum.
The spectrum of 2MJ1237 appears to be a single star with an RV of 6.2$\pm$1.0 km s$^{-1}$ and $v$sin$i$\textgreater 35 km s$^{-1}$. 
This RV for 2MJ1237 is inconsistent with the value of 11.3$\pm$2.4 km s$^{-1}$ measured by \cite{riedel2016} using low-resolution optical spectra, but is still consistent with the RV distribution of TWA Members. The optimal RV given by BANYAN $\Sigma$, assuming that 2MASS~J1237 is a member of TWA, is 7.7$\pm$1.5 km s$^{-1}$, which is consistent with our measured RV. Differing radial velocities may be the result of unresolved binarity in the regime of a small flux ratio and small total velocity separations, or \cite{riedel2016} may have underestimated the uncertainties they obtained using R$=$5000 spectra on an object with large $v$sin$i$. Using BANYAN $\Sigma$ and our measured RV, we find that 2MJ1237 has a TWA membership probability of 99.5$\%$. 

\section{Discussion}
The identification of 2MJ1235AB as a double-lined spectroscopic binary, with a system radial velocity of -6.47$\pm$3.9 km s$^{-1}$, makes it a debatable member of TWA. 
However, we can not say that 2MJ1235AB is not a member of TWA, despite the low membership probability given by Banyan $\Sigma$. 
The high-quality of the match of TWA~9B to both components of 2MJ1235AB simply based on visual inspection (Figure~\ref{binary_fit}) led us to question whether or not the goodness-of-fit when using a TWA member as a template might be a youth indicator for 2MJ1235AB in addition to the age range we determined by placing 2MJ1235AB on the H-R diagram. To explore the possibility of templates with older ages providing a better or worse template fit than the $\sim$10~Myr TWA members, we performed the same routine we used to fit veiling, RV, and flux ratio for 2MJ1235AB but with the older field stars GU Psc and Wolf 1130 in place of TWA 9B.
GU Psc is a member of the AB Doradus YMG, an association that has an age of $\sim$100~Myr \citep{luhman2005}. Wolf 1130 is a field star with an age \textgreater 3.7 Gyr \citep{mace2018}. 
The chi-squared values for the template fits across all IGRINS orders using TWA 9B, GU Psc and Wolf 1130 are 0.6, 1.3 and 1.2, respectively.
The good match between TWA 9B and 2MJ1235AB is based on a complicated combination of physical properties like temperature, surface gravity and veiling.
The accuracy of the cross-correlation that we perform is dominated by spectral line locations, which is primarily sensitive to temperature and RV. 
\citet{Kraus2011} discuss the inflated radii of close binaries like 2MJ1235AB, and we would expect surface gravity to be lower when the radius is enlarged.
The effects of surface gravity and continuum veiling on spectral line depths are not easily disentangled, but has little effect on measured radial velocities.
Based on this analysis, ages $>$100~Myr are inconsistent with the spectra of 2MJ1235AB.

\begin{figure}[ht]
\centering
\includegraphics[width=6.5in]{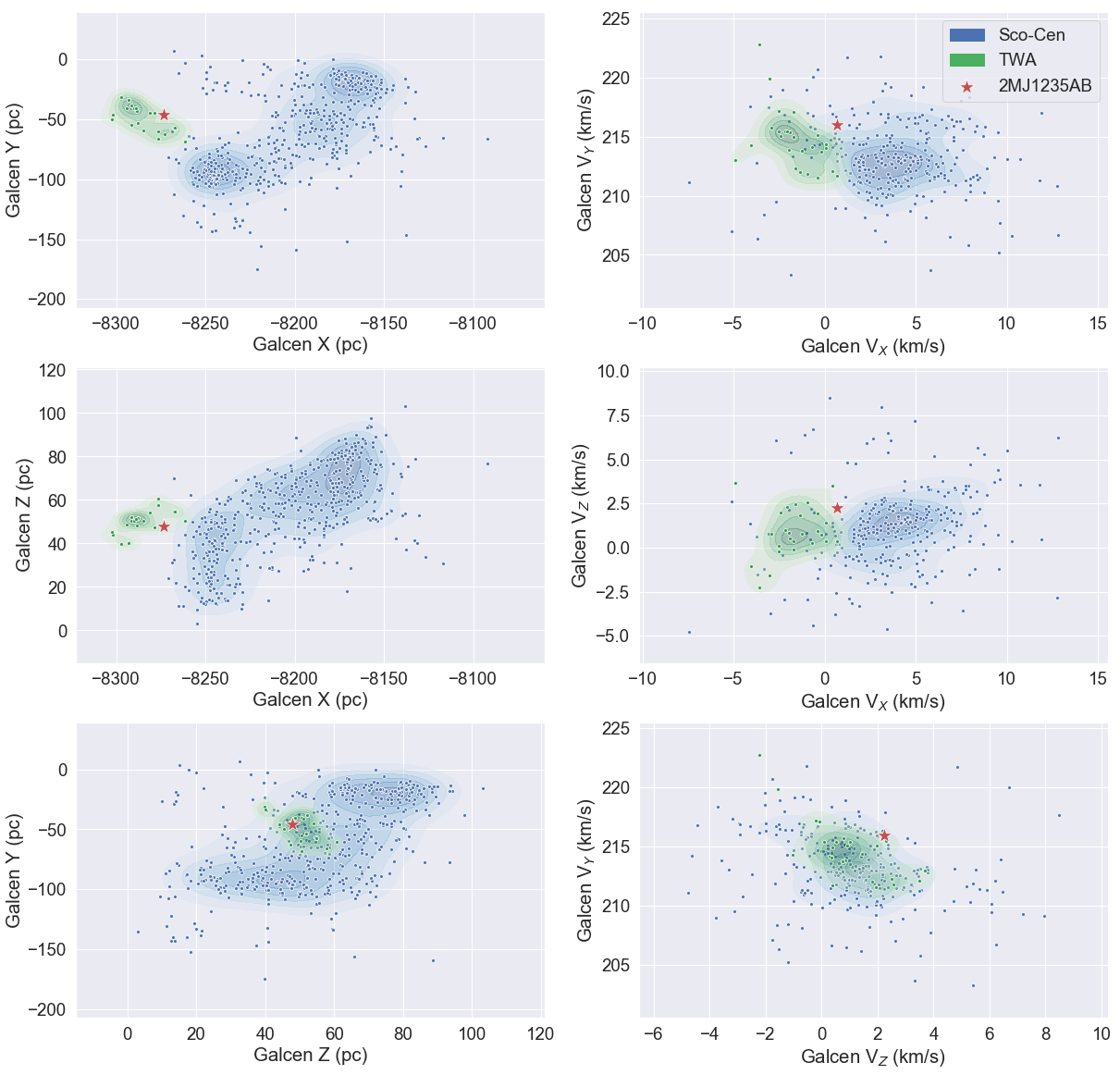}

\caption{ Galactocentric positions and velocities for 2MJ1235AB in relation to members of TWA and Sco-Cen. The figures in the left column show galactocentric positions and those in the right column show galactocentric velocities. The individual positions and velocities of members are shown as well as contours showing a kernel density estimation in each plane for TWA and Sco-Cen. Membership information for Sco-Cen was taken from \citet{pecaut2016}.  \label{scocen_twa_pos_vel}}
\end{figure}

To examine the space position and velocity characteristics of 2MJ1235AB with respect to TWA, in Figure \ref{scocen_twa_pos_vel} we show the galactocentric XYZ positions and velocities along with other members of TWA and the nearby Scorpius-Centaurus (Sco-Cen) OB Association \citep{pecaut2016}. 
The XYZ position of 2MJ1235AB falls on the outskirts of the spatial extent of bona fide members of TWA as determined by Gaussian kernel density estimations shown in Figure \ref{scocen_twa_pos_vel}. 
However, based on position alone, 2MJ1235AB cannot be ruled out as an outlying member of Sco-Cen. As for the position of 2MJ1235AB in velocity space, the system falls in the overlap region of TWA and Sco-Cen. While the position and velocity of the system do not provide definitive evidence for membership of the system in either TWA or Sco-Cen, it may be possible to determine membership based on age and/or traceback. 2MJ1235AB was chosen as a candidate member of TWA partially because the presence of H$\alpha$ emission and its X-ray luminosity \citep{riaz2006}. 
Yet, binarity can prolong stellar activity, and close binaries can maintain their activity for Gyrs \citep{mace2018}. In short, close binarity makes it difficult to determine age accurately enough to group 2MJ1235AB with the $\sim$8 Myr TWA or the $\sim$16 Myr \citep{mamajek2002} Lower-Centaurus Crux, a sub-group of Sco-Cen. We also note that the approximate age of the Lower-Centaurus Crux falls within the age range we determined for 2MJ1235AB of 12-50~Myr \cite[and the age estimated by ][]{donaldson2016}, but the age for TWA falls slightly below this range.  While some bounds could be placed on the age of 2MJ1235 using gravity sensitive spectral indices \citep{allers2013, martin2017}, the strength of lithium absorption at these spectral types would discern between the 10~Myr and 50~Myr ages \citep{Mentuch2008}.

We have also explored the possibility of spatially resolving this binary pair. However, our measured upper limit on the separation translates to an angular separation of 2.2 mas on the sky. The largest telescopes currently in operation can only achieve angular resolution $\sim$20 times larger than what would be needed to resolve this pair. For example, the W.M. Keck telescopes can achieve angular resolution of 40 mas with adaptive optics \citep{wizinowich2000}. Planned extremely large telescopes such as the Giant Magellan Telescope will improve this angular resolution by a factor of 4 \citep{johns2012}. However, this is still $\sim$5 times larger than what would be needed to potentially resolve 2MJ1235AB. The predicted capabilities of {\it JWST} are also an order of magnitude greater than would be required for this system \citep{perrin2014}.

In the future, traceback analysis may be a better route for determining membership and age information on potential YMG membership. \cite{donaldson2016} performed a traceback analysis of core members of TWA that did not result in convergence. \cite{donaldson2016} attributed this result to the inaccuracy of current RV measurements of the core members of TWA and stated that a precision of 0.25 km $s^{-1}$ is needed in order to produce accurate traceback information for bona fide and potential members. Our own determination of the center of mass velocity for 2MJ1235AB has an uncertainty that is $\sim$8 times the level needed to produce useful information from traceback analysis. 
This would be improved through an observing campaign to derive the full orbital solution of 2MJ1235AB, but the primary source of RV uncertainty is in converting relative RVs to absolute RVs through the use of RV standards \citep{mace2016}.
Given the escalating development of precision RV instrumentation for exoplanet searches, determining membership based on tracebacks for systems such as 2MJ1235AB may become possible in the next several years. 

 2MJ1235AB reveals some of the limitations of kinematic surveys searching for members of YMGs and it may be an older system masquerading as a younger one due to prolonged activity caused by its binarity.
 It may also be a pair of field stars that formed in a larger molecular complex connected to TWA and other nearby YMGs.
 \cite{gagne2018b} found preliminary evidence that some YMGs, including TWA, may be spatially extended to larger distances than previously thought, and that some of these spatial outliers may form a bridge between TWA and the Lower Centaurus Crux. This new evidence implies that YMGs may not only be more spatially extended, but more interconnected than previously thought. 
 In the case of 2MJ1235AB, we find no direct evidence that the system could be linked to any other YMGs in Banyan $\Sigma$.
 Determining YMG membership is useful for studying young stellar populations, but it is also important to recognize that systems like 2MJ1235AB may play an important roll in tracing back the origins of YMGs and young field stars alike. 
 We emphasize that based on age indicators, astrometry, and projected kinematics - 2MJ1235AB fully appeared to be a member of TWA, but high-resolution spectroscopic analysis quickly revealed its binarity and outlier kinematics. 
 We therefore suggest that more robust RV measurements and spectroscopic analysis be performed before giving objects names associated with specific YMGs. 

\software{IGRINS pipeline (Lee \& Gullikson 2016), ZBARYCORR (Wright \& Eastman 2014), BANYAN Σ (Gagne et al. 2018)}

\FloatBarrier

\acknowledgements

We thank the anonymous referee who provided detailed and insightful comments that improved the robustness of the results presented here as well as the clarity of the text. This work used the Immersion Grating Infrared Spectrometer (IGRINS) that was developed under a collaboration between the University of Texas at Austin and the Korea Astronomy and Space Science Institute (KASI) with the financial support of the US National Science Foundation under grants AST-1229522 and AST-1702267, of the University of Texas at Austin, and of the Korean GMT Project of KASI.
Based on observations obtained at the Gemini Observatory, which is operated by the Association of Universities for Research in Astronomy, Inc., under a cooperative agreement with the NSF on behalf of the Gemini partnership: the National Science Foundation (United States), the National Research Council (Canada), CONICYT (Chile), Ministerio de Ciencia, Tecnolog\'{i}a e Innovaci\'{o}n Productiva (Argentina), and Minist\'{e}rio da Ci\^{e}ncia, Tecnologia e Inova\c{c}\~{a}o (Brazil).
This work has made use of data from the European Space Agency (ESA)
mission {\it Gaia} (\url{https://www.cosmos.esa.int/gaia}), processed by
the {\it Gaia} Data Processing and Analysis Consortium (DPAC,
\url{https://www.cosmos.esa.int/web/gaia/dpac/consortium}). Funding
for the DPAC has been provided by national institutions, in particular
the institutions participating in the {\it Gaia} Multilateral Agreement.

\newpage

\bibliography{twa46_letter.bib}

\newpage

\end{document}